\documentclass[aps,prl,showpacs,twocolumn]{revtex4-1}
\usepackage{amsmath,amssymb,graphicx}

\newcommand{\be}{\begin{equation}}
\newcommand{\ee}{\end{equation}}
\newcommand{\bea}{\begin{eqnarray}}
\newcommand{\eea}{\end{eqnarray}}

\begin{document}

\noindent{\bf Comment on ``Dark Matter with Pseudoscalar-Mediated Interactions Explains the DAMA Signal and the Galactic Center Excess''}\\


\noindent In a recent Letter  \cite{Arina:2014yna}, the Dirac fermionic dark matter with pseudoscalar-mediated interactions has been proposed as an explanation for the Galactic Center excess, correct relic density and DAMA signal.
The DAMA regions may be rejected by flavor constraints \cite{Dolan:2014ska}. However, since the process is approximated by one-particle exchange in the simplified model approach, the inclusion of multi-particle exchange processes may change the form of  WIMP-nucleus cross sections, and all bounds shall be reconsidered \cite{DelNobile:2015lxa}.
 The authors of \cite{Arina:2014yna} assumed that contact interactions, with ${g_{\rm DM} g}/({|\vec{q}|^2 +m_a^2}$) replaced by $1/\Lambda_a^2$,  remain roughly valid in calculating scattering rates at the direct detection even when the mediator mass $m_a$ is the same order as the typical momentum transfer $|\vec{q}|$. 
To account for these observables,  they find that $m_a$ is of order 50 MeV. They expect small changes occur in their fit to DAMA due to onset of the long-range region but conclusions will not be modified.
 
We will show that such a replacement is not suitable. We follow  \cite{Arina:2014yna} for notations and inputs. Adopting full form of interactions, Figs.~1(a) and 2(a) show DAMA results for universal and Higgs-like couplings in the $m_{\rm DM}$-$m_a/\sqrt{g_{\rm DM} g}$ plane with respect to different values of $m_a$. $m_a$ dependence of $m_a/\sqrt{g_{\rm DM} g}$ is thus shown. 
 
\begin{figure}[t!]
  \begin{center}
 \includegraphics[width=0.39\textwidth]{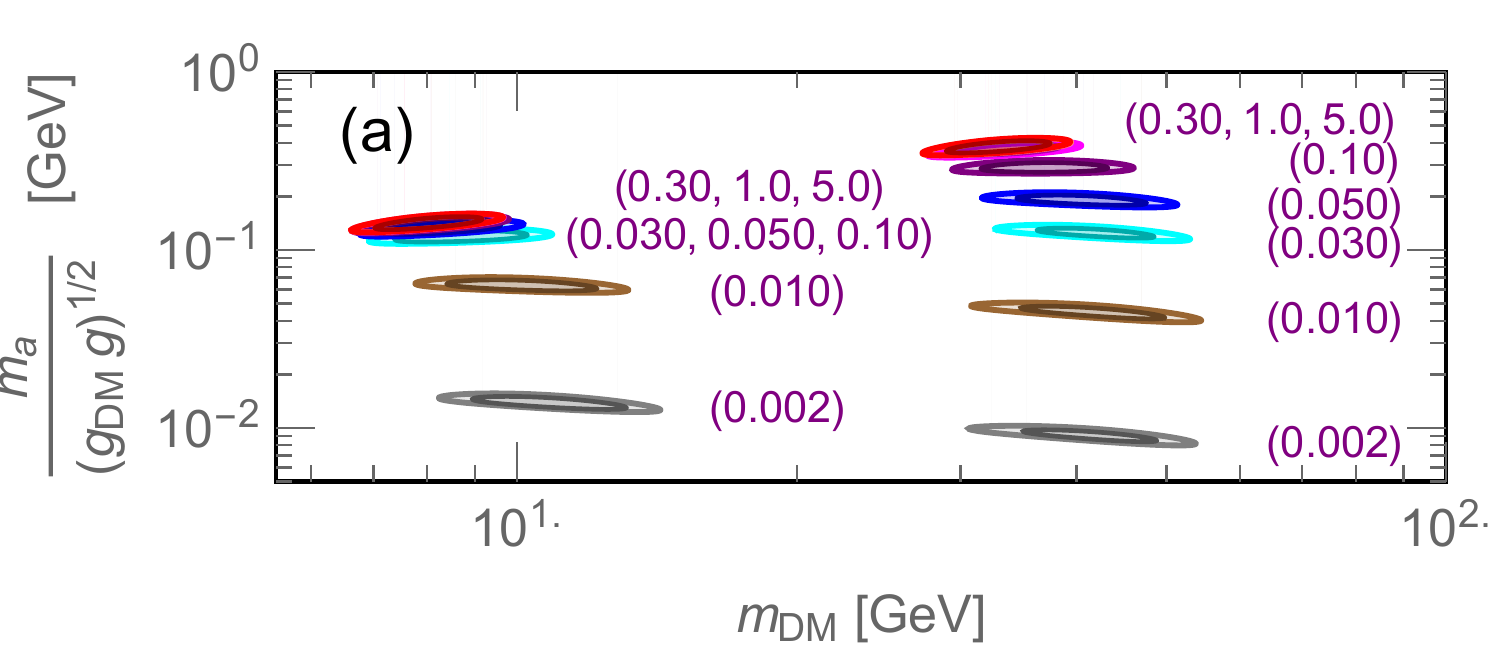}
 \\
\includegraphics[width=0.39\textwidth]{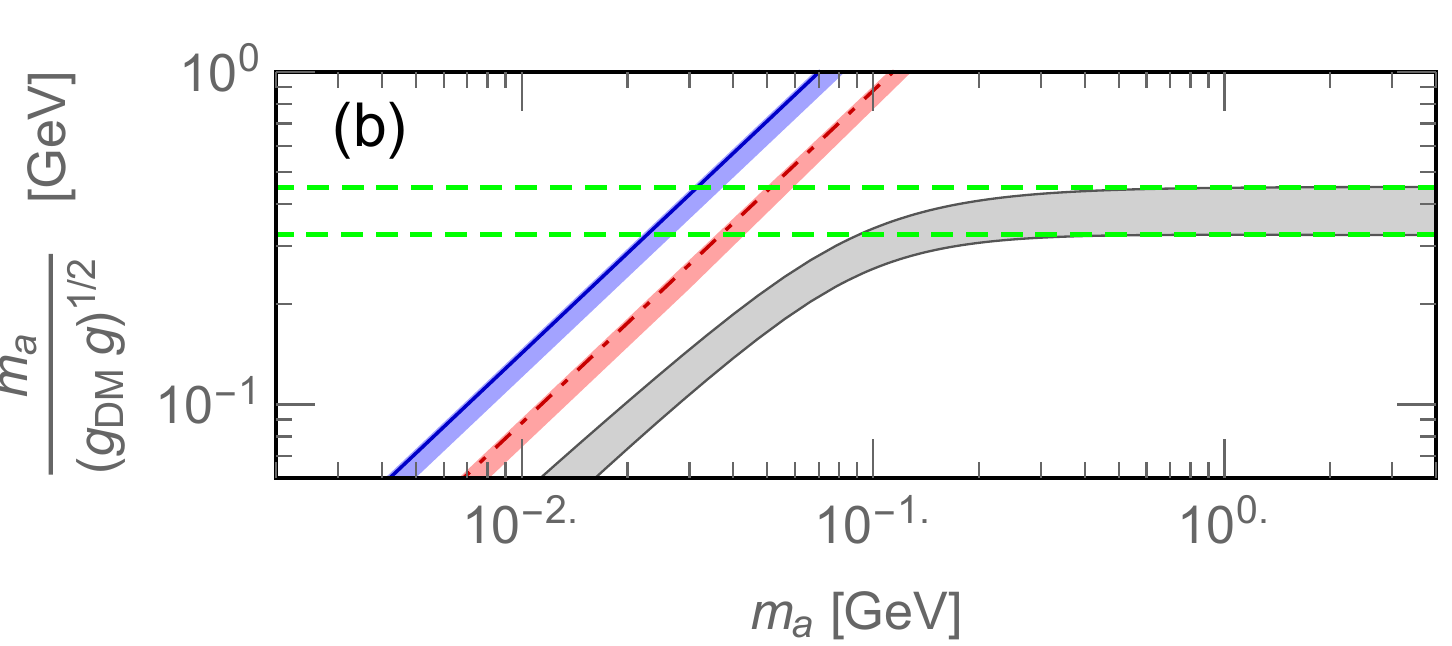} 
\caption{\label{fig:quc} (a)  2$\sigma$ (inner) and 3$\sigma$ (outer) DAMA regions in the $( m_{\rm DM}, m_a/\sqrt{g_{\rm DM}g})$ plane for (heavy-)flavor-universal couplings. The corresponding number in parentheses is $m_a$ in units of GeV.  (b) 3$\sigma$ allowed  regions for DAMA (gray) and for $\gamma$-ray excess with flavor-universal (blue) and heavy-flavor-universal couplings (red). The range between dashed lines is for the contact limit.}
\end{center}
\end{figure}

\begin{figure}[t!]
\includegraphics[width=0.38\textwidth]{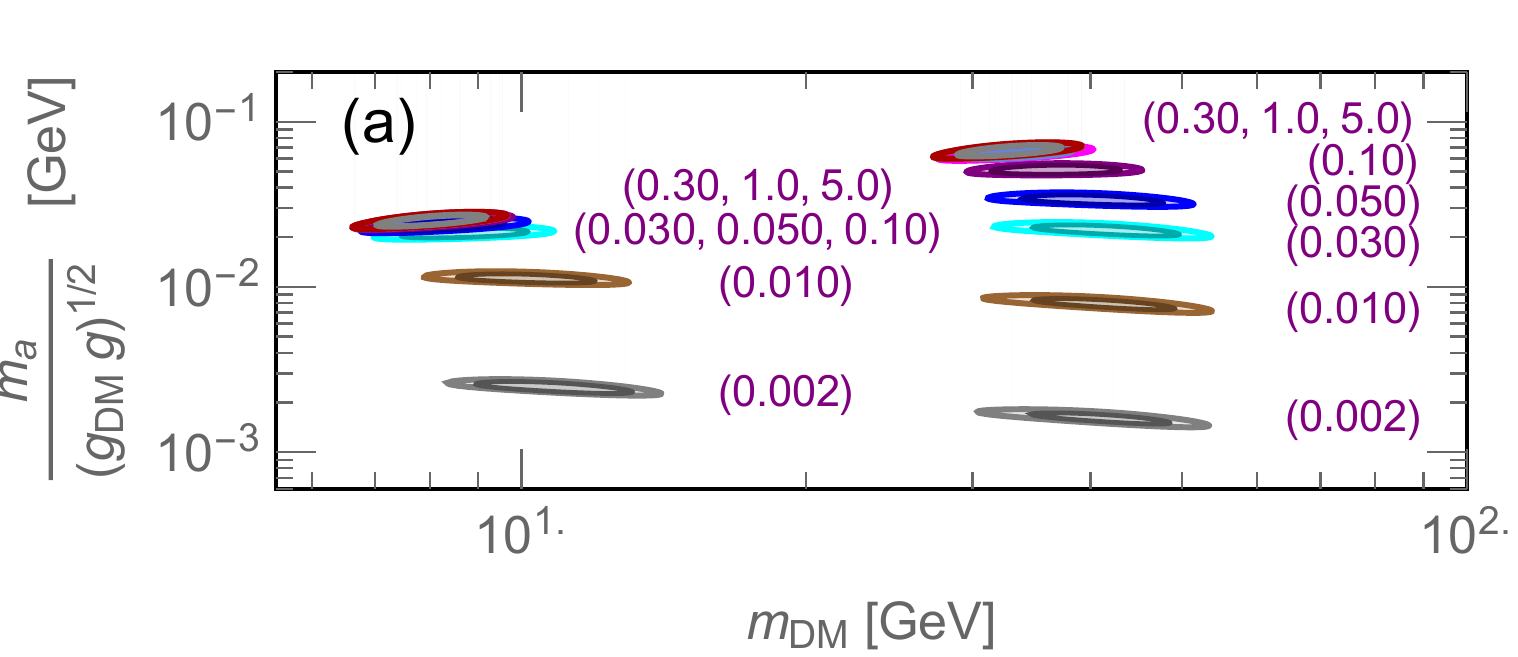} 
\\
\includegraphics[width=0.38\textwidth]{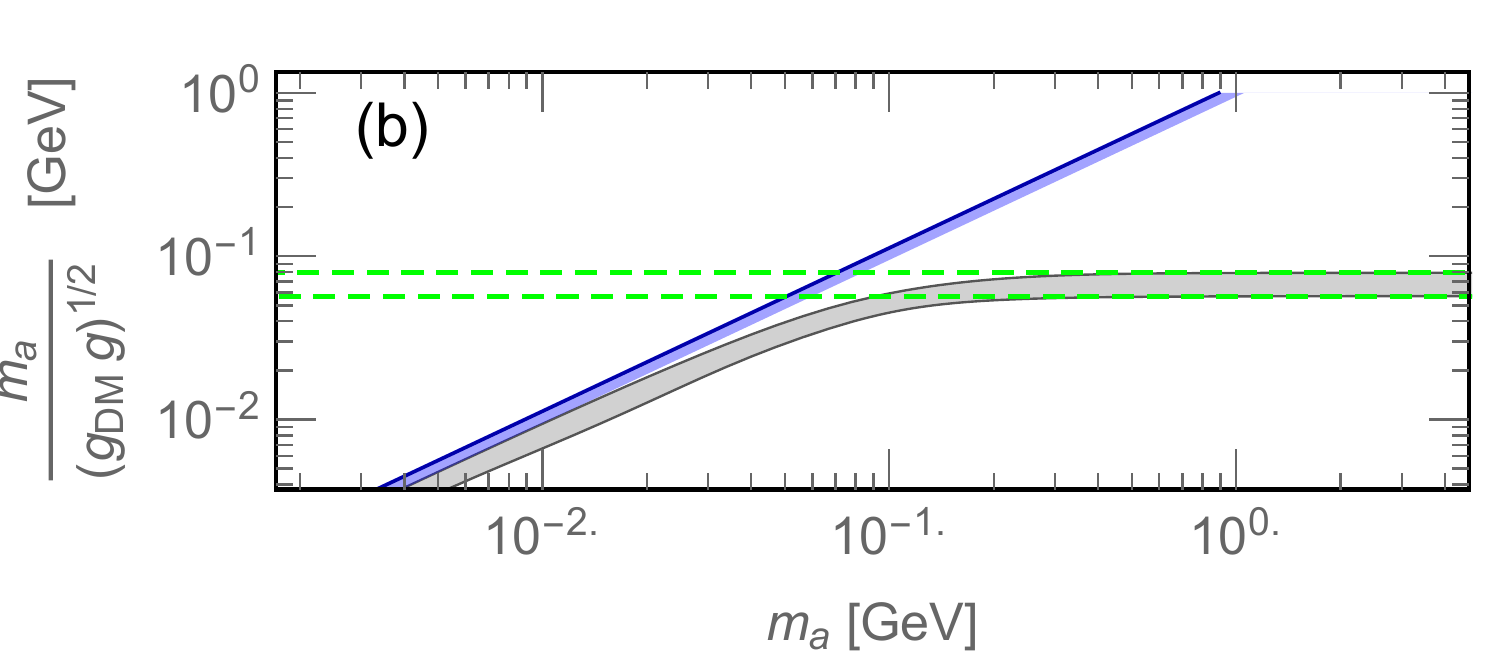} 
\caption{\label{fig:higgs} Same as Fig. \ref{fig:quc} except for Higgs-like couplings.}
\end{figure}

In Figs.~\ref{fig:quc}(b) and \ref{fig:higgs}(b), we plot DAMA regions with dark matter mass $m_{\rm DM}\sim40$ GeV, further including uncertainties due to variations of quark masses and $\Delta{q}^{(N)}$, and $\gamma$-ray excess regions in the $m_{a}$-$m_a/\sqrt{g_{\rm DM} g}$ plane. The DAMA (gray shaded) regions  satisfy that $m_a/\sqrt{g_{\rm DM} g}$ (or  $g_{\rm DM}g$) is constant for $m_a$ $\gg$ (or $\ll$) 100 MeV.  In Fig.~\ref{fig:quc}(b), the solid and dotdashed lines,  corresponding to $\gamma$-ray excess best fits given in \cite{Arina:2014yna}, are for universal couplings to all quarks and only to heavy quarks ($c, b, t$), respectively, where shaded regions are uncertainties (3$\sigma$) from updated analyses \cite{Daylan:2014rsa}. It shows that DAMA and $\gamma$-ray excess regions do not overlap
, even for heavy-flavor-universal couplings, for which  $m_{\rm DM}\approx30\sim50$~GeV in the $\gamma$-ray excess fit. 
(It was found in \cite{Dolan:2014ska} that for flavor universal couplings, the DAMA signal is incompatible with the thermal relic requirement.)
As for Higgs-like couplings, the two regions overlap for $m_a\lesssim$ 15 MeV, where long-range interactions, instead of contact interactions, occur at the DAMA.


 In summary, we did not find the conclusion in \cite{Arina:2014yna} warranted.

\vspace{0mm}

\noindent {\it Acknowledgement.} 
This work was supported in part by the Ministry of Science and Technology of R.O.C. under Grant No: 102-2112-M-033-007-MY3.
\\
\vskip30mm
\noindent{Kwei-Chou Yang}\\
{\small Department of Physics and Center for High Energy Physics, Chung Yuan Christian University, Taoyuan 320, Taiwan}

\vskip0mm

\vspace{3mm}


\end{document}